\documentclass[letterpaper, 10 pt, conference]{ieeeconf}  

\IEEEoverridecommandlockouts                              

\usepackage{graphics} 
\usepackage{epsfig} 
\usepackage{mathrsfs}
\usepackage{times} 
\usepackage{amsmath} 
\usepackage{amssymb}  
\usepackage{amsfonts}
\usepackage{amssymb}
\usepackage{pgf}
\usepackage{graphicx}
\usepackage{mathrsfs}
\usepackage{enumerate}
\usepackage{soul}
\usepackage{tikz}
\usepackage{pgfplots}
\usepackage{standalone}
\usepackage{subfigure}
\usepackage{caption}
\pgfplotsset{width=9cm,height=4.5cm,compat=1.16}
\pgfplotsset{every tick label/.append style={font=\tiny}}
\pgfplotsset{tick scale binop=\times}
\usepackage{mathtools}
\usetikzlibrary{arrows}
\usepackage[noadjust]{cite}
\usepackage[utf8]{inputenc}
\usepackage{pgfplots}
\pgfplotsset{compat=newest}
\usepgfplotslibrary{groupplots}
\usepgfplotslibrary{dateplot}
\usepackage{url}
\newcommand{\Var}{\mathrm{var}}

\usepackage[T1]{fontenc}
\usepackage{upgreek}
\usepackage{subfig} 
\usepackage{bm}
\usepackage{color}
\newcommand{\AY}[1]{\textcolor{black}{#1}}

\usepackage{algorithm,algorithmicx}
\usepackage{algpseudocode}

\DeclareMathOperator*{\argmin}{arg\,min}

\DeclareFontFamily{U}{matha}{\hyphenchar\font45}
\DeclareFontShape{U}{matha}{m}{n}{
      <5> <6> <7> <8> <9> <10> gen * matha
      <10.95> matha10 <12> <14.4> <17.28> <20.74> <24.88> matha12
      }{}
\DeclareSymbolFont{matha}{U}{matha}{m}{n}
\DeclareMathSymbol{\varsubseteq}{3}{matha}{"84}
\DeclareMathSymbol{\in}{3}{matha}{"50}

\DeclareFontEncoding{LS1}{}{}
\DeclareFontSubstitution{LS1}{stix}{m}{n}
\DeclareSymbolFont{symbols2}{LS1}{stixfrak}{m}{n}
\DeclareSymbolFont{symbols3}{LS1}{stixbb}{m}{n}
\SetSymbolFont{symbols2}{bold}{LS1}{stixfrak}{b}{n}
\SetSymbolFont{symbols3}{bold}{LS1}{stixbb}{b}{n}
\DeclareSymbolFont{arrows1}{LS1}{stixsf}{m}{n}
\SetSymbolFont{arrows1}{bold}{LS1}{stixsf}{b}{n}
\DeclareMathSymbol{\varrightarrow}{\mathrel}{arrows1}{"99}
\DeclareMathSymbol{\varleftarrow}{\mathrel}{arrows1}{"7D}
\DeclareMathSymbol{\fourvdots}{\mathord}{symbols2}{"38}

\newtheorem{thm}{Theorem}[section]
\newtheorem{lem}[thm]{Lemma}

\newtheorem{rem}{Remark}

\newtheorem{assump}{Assumption}

\title{\LARGE \bf
A Model-Based Reinforcement Learning Approach for PID Design
}

\author{Hozefa Jesawada, Amol Yerudkar, Carmen Del Vecchio, and Navdeep Singh
\thanks{H. Jesawada, A. Yerudkar and C. Del Vecchio are with the Department of Engineering, University of Sannio, Benevento 82100, Italy.
        {\tt\small \{jesawada,\ ayerudkar,\ c.delvecchio\}@unisannio.it}.}%
\thanks{N. Singh is with the Department of Electrical Engineering, Veermata Jijabai Technological Institute, Mumbai 400019, India.
       {\tt\small \{nmsingh\}@ee.vjti.ac.in}.}        
}

\begin{document}
\maketitle
\thispagestyle{empty}
\pagestyle{empty}

\begin{abstract} 
Proportional-integral-derivative (PID) controller is widely used across various industrial process control applications because of its straightforward implementation. However, it can be challenging to fine-tune the PID parameters in practice to achieve robust performance. \AY{The paper proposes a model-based reinforcement learning (RL) framework to design PID controllers leveraging the probabilistic inference for learning control (PILCO) method and Kullback-Leibler divergence (KLD). Since PID controllers have a much more interpretable control structure than a network basis function, an optimal policy given by PILCO is transformed into a set of robust PID tuning parameters for underactuated mechanical systems. The presented method is general and can blend with several model-based and model-free algorithms.} The performance of the devised PID controllers is demonstrated with simulation studies for a benchmark cart-pole system under disturbances and system parameter uncertainties. 
\end{abstract}
\section{Introduction}
Proportional-integral-derivative (PID) controllers overwhelmingly dominate wide range of industrial processes owing to their simplicity and efficiency~\cite{aastrom1995pid}. Designing a control scheme necessitates tuning the PID gains, which can be done manually or through a heuristic method. For example, Ziegler-Nichols~\cite{ziegler1942optimum} is a well-known method for achieving the necessary performance for uncertain systems. This method is challenging to implement and can make the system susceptible to parameter changes, thereby failing to maintain control performance. The dominant pole placement and biggest log modulus methods are also used widely for multiple-input-multiple-output (MIMO) systems~\cite{o2009handbook}. Such approaches aim to fine-tune each control loop independently. In the last decades, many researchers have proposed adaptive PID control strategies, for instance, see~\cite{yu2007stable,yamamoto2004design} and the references therein. However, these tuning methods resort to model-based control theory, wherein firstly mathematical model is constructed using the first principles or system identification which then is utilized to devise the controllers. It is observed that the derived model only approximates the actual dynamical system with significant uncertainties, which raises a number of practical issues.  

Reinforcement learning (RL) has spurred widespread interest in the research community by introducing data-driven methods as an alternative to model-based control techniques. Authors in~\cite{khong2016iterative} and~\cite{zhang2017data} have presented data-driven PID tuning techniques based on iterative learning control and one-shot experiment data, respectively. In addition, some modern data-driven PID tuning methods such as deep learning, RL and genetic optimization have been presented in~\cite{carlucho2020adaptive,jaen2013pid}. 
A tuning method by employing RL and deep neural networks has been derived in~\cite{shipman2019reinforcement}. \AY{Bayesian optimization and Gaussian process (GP) methods have recently been investigated for tuning controllers without previous knowledge of the system dynamics~\cite{marco2016automatic,neumann2019data}.} Doerr \textit{et al.}~\cite{doerr2017model} proposed a model-based RL framework by resorting to one of the famous model-based RL approaches, namely, probabilistic inference for learning control
(PILCO)~\cite{deisenroth2011pilco}, which has a data-efficient structure that relies on Bayesian inference with the GP. In particular, authors in~\cite{doerr2017model} obtained optimal PID parameters for a seven degree-of-freedom robot arm balancing an inverted pendulum by considering the PID as an RL agent with augmented state-space and analytically solving error gradients to design the optimal gains. \AY{Recently, Yoo \textit{et al.}~\cite{yoo2020hybrid} integrated the linear combination of deterministic controllers (a PD controller and an LQR) and a parameterized policy directly into the PILCO algorithm to achieve faster convergence. To the best of the authors' knowledge, only~\cite{doerr2017model} deals with optimizing PID gains for MIMO systems by modifying the PILCO. However, there are some shortcomings of PILCO, such as limited handling of low-data regimes, a priori parameterized policy requirements, no possibility to include safety constraints, etc. Several algorithms have been presented to overcome PILCO's limitations. For example, DeepPILCO~\cite{gal2016improving}, safePILCO~\cite{polymenakos2020safe} and model-based policy optimization (MBPO)~\cite{janner2019trust}. Nonetheless, the results in~\cite{doerr2017model} cannot be extended to the algorithms in~\cite{gal2016improving,polymenakos2020safe,janner2019trust}.} Thus, designing a structurally simple, efficient, and robust PID controller using a model-based RL framework deserves more attention from the research community. 

By following this stream of research, \AY{in this paper, we present a generalized method for designing PID controllers that can hinge on several model-based and model-free RL algorithms. To demonstrate the results, we select a benchmark data-efficient PILCO algorithm and exploit its inherent properties to develop robust PID controllers for a general class of systems.} Different from~\cite{doerr2017model} and~\cite{yoo2020hybrid}, in this paper, we utilize Kullback-Leibler divergence (KLD)~\cite{kullback1951information} to transform the PILCO policy into a interpretable control design, i.e., a PID controller for underactuated systems. The designed controller is robust against disturbances and system parameter uncertainties. The main contributions of this paper as follows: \AY{i)~We propose KLD-based generalized framework to design optimal PID tuning parameters for MIMO systems. The designed controllers utilize recorded system data and are iteratively optimized without a priori knowledge of the system dynamics; ii)~We leverage the PILCO structure to present the main results on PID design. The proposed method is demonstrated by a simulation study on an underactuated cart-pole system. The designed PID controller shows robust behavior when tested under matched and unmatched disturbances and some system parameter uncertainties; iii)~To illustrate the versatility of our method, we also design PID controllers using MBPO as the underlying structure instead of PILCO and show the closed-loop performance of the cart-pole system; iv)~A closed-loop stability study is performed with region of attraction (ROA) analysis to show the viability of the presented results.}

\section{Preliminaries and Problem Formulation}\label{Prelimnaries}
This section introduces concepts of PID control scheme and model-based RL followed by the problem formulation.

Consider a discrete dynamical system with state $x_{\tau}\in \mathbb{R}^{M}$ and input $u_{\tau}\in \mathbb{R}^{N}$, represented as
\begin{equation}\label{model}
    x_{\tau+1}=g(x_{\tau},u_{\tau})+\omega_{\tau},
\end{equation}
where $g(.)$ is some nonlinear function defining unknown system dynamics and $\omega_{\tau}$ is the output uncertainty component derived as the the normal distribution $\mathcal{N}\left ( 0,\Sigma_{\omega} \right )$. The objective is to design a robust PID controller
that renders the system asymptotically stable under input disturbances and system parameter variations.  
\subsection{PID controller}
PID controller is the most common type of controller used in practice due to its typical structure, high reliability, and ease of operation. The basic structure of the PID controller is defined as follows $u_{\tau} = K_{P}e_{\tau}+K_{I}\int_{0}^{\tau}{e_{\tau}}d\tau+K_{D}\frac{de_{\tau}}{d\tau},$
where $u_{\tau}$ is the control input of a plant, $e_\tau$ is the error term derived from desired output denoted by $x_{des}$ and plant output $x_\tau$ as $e_\tau=x_{des}-x_{\tau},$
$K_{P}$, $K_{I}$, and $K_{D}$ denote gains of proportional, integral, and derivative terms. There are various methods for tuning controller gains in literature~\cite{cominos2002pid}. Though controllers for the lower-dimensional systems can be tuned manually, tuning controller gains for higher dimension systems with under-actuated dynamics is a difficult task.

Underactuated systems are characterized by
the fact that there are more degrees of freedom than actuators, i.e., one or more degrees of freedom are unactuated. 
The underactuated dynamics may cause controller couplings, and tuning each controller independently may not be feasible. The proposed method can overcome such challenges posed by the coupling of the controllers.
The possible PID structures are exemplified with two controllers as shown below in \eqref{PID_2} and \eqref{PID_1}.
\begin{equation}\label{PID_2}
    K=\begin{bmatrix}
K_{p1} & 0 & K_{i1} & 0 & K_{d1} & 0\\
0 & K_{p2} & 0 & K_{i2} & 0 & K_{d2}\\
\end{bmatrix},
\end{equation}
 \begin{equation}\label{PID_1}
    K=\begin{bmatrix}
K_{p1} & K_{p2} & K_{i1} & K_{i2} & K_{d1} & K_{d2}
\end{bmatrix}.
\end{equation}
The two controllers acting on different input is shown in~\eqref{PID_2}, while~\eqref{PID_1} represents the additive action on single input~\cite{doerr2017model}. We present a generalized data-driven RL-based method that can deal with the aforementioned PID structures.


\subsection{Reinforcement Learning: A Model-Based Approach}\label{PILCO_sec}
RL is one of the paradigms of machine learning which focuses on learning system dynamics through interactions with the environment~\cite{sutton2018reinforcement}. RL algorithms are broadly classified as model-free and model-based. This paper deals with the model-based RL algorithms.  

Model-based RL methods are more promising in terms of data-efficiency while learning control policies \cite{deisenroth2011pilco,neumann2019data}. One such model-based on-policy RL method which has gain popularity is PILCO~\cite{deisenroth2011pilco}. This method learns the dynamics model using Bayesian inference with GP, predicts the cost using the learned dynamics and uncertainty propagation, and then performs gradient descent to find the optimal policy.

The objective of PILCO is to find a policy $\pi_\varphi$ controlling a system towards a desired state, parameterized by $\varphi$
that minimizes the expected cost-to-go $J_{\pi}(\varphi)$ derived as 
\begin{equation}
J_{\pi}(\varphi)=  \sum_{\tau=0}^{T}\mathbb{E}_{x_{\tau}}[c(x_{\tau})], \quad x_{0}\sim\mathcal{N}(\mu_{0},\Sigma_{0}),
\end{equation}
where $c(x_{\tau})$ is the immediate cost of being in state $x$ at time $\tau$ w.r.t. to the desired state $x_{des}$, and $\mathbb{E}[\,\cdot\,]$ is the expected value operator. The initial state $x_{0}$ is considered to be random function with mean ($\mu_{0}$) and variance ($\Sigma_{0}$).

PILCO employs GP to model the system dynamics denoted as $\hat{g}(\cdot)$, where tuple $(x_{\tau},u_{\tau})\in \mathbb{R}^{M+N}$ is considered model input and differences $\Delta_{\tau}=x_{\tau+1}-x_{\tau} \in~\mathbb{R}^{M}$ as training targets. The joint probability density function (pdf) of $x_{\tau+1}$ for a given $(x_{\tau},u_{\tau})$ is derived as
\begin{equation}\label{e1}
     \textit{p}(x_{\tau+1}\mid x_{\tau},u_{\tau})=\mathcal{N}(x_{\tau+1}\mid\mu_{\tau+1}, \Sigma_{\tau+1}),
 \end{equation}
 where the mean function is $\mu_{\tau+1}=x_{\tau}+\mathbb{E}_{g}[\Delta_{\tau}],$ and variance function is $\Sigma_{\tau+1}=\Var_g[\Delta_{\tau}].$

The GP output is prediction with $[\tilde{x}]:=[x^{\top} u^{\top}]^{\top}$. The training input and target data matrix for $n$ samples are represented as $\tilde{X}=[\tilde{x}_{1},\dots,\tilde{x}_{n}]$, and $Y =[\Delta_{1},\dots,\Delta_{n}]^{\top}$, respectively. The data $Y\sim \mathcal{N}(0, \mathbb{K})$ with $\mathbb{K}=\Sigma_{\eta}+\Sigma_{\omega}^2I$,
and $\mathrm{M\times M}$ identity matrix $I$, is represented in the Bayesian approach. The elements of $\Sigma_{\eta}$ are rewritten by $\Sigma_{ij}=\rm{cov}(\Delta_{i}, \Delta_{j})=\mathbb{C}(\tilde{x}_i,\tilde{x}_j),$
%
 %
where $\mathbb{C}(\cdot,\,\cdot)$ is a positive semidefinite covariance function, which is also called \textit{kernel}.  We refer to~\cite{rasmussen2003gaussian} for a detailed discussion on GP-based policy search.

\AY{In PILCO the policy is parameterized either as a network of Gaussian/squared exponential basis functions or as a linear policy.} This policy evaluation loop is terminated once the optimal policy is learned, i.e., the optimum parameters $\theta^{*}$ are attained. On learning a optimal policy, it is rolled-out to the system and a new data set is generated. 
Combined with previously generated data, the data-set is employed to refine and update the dynamical model of the system.

\subsection{Problem formulation: learning PID gains through $\pi_{\varphi}$}

The model-based RL method namely PILCO \AY{(originally introduced in~\cite{deisenroth2011pilco})} described in Section~\ref{PILCO_sec} is summarized in the following algorithm.
\begin{algorithm}[h]\label{ALgo_1}
\caption{Model-Based Optimal Policy Search} 
\begin{algorithmic}[1]
\State{Initialize parameters of the controller by sampling distribution $\varphi\sim\mathcal{N}(0,\sigma)$. Perform random policy roll-out and collect data.}

\State{\algorithmicrepeat

Learn system dynamics using all the collected data.

\algorithmicrepeat

\begin{itemize}
    \item $\quad$ Roll-out: Compute $J_{\pi}(\varphi)$ given $\pi_{\varphi}(\tilde{X}_{\tau},\varphi)$ and $\hat{g}(\cdot)$
    \item $\quad$ Analytically compute policy gradient $dJ/d\varphi.$
    \item $\quad$ Gradient based update of parameters $\varphi$.
\end{itemize}

\algorithmicuntil \hspace{0.1cm} Convergence to $\varphi^*$.}

\AY{Perform policy $\pi_{\varphi}^*$ roll-out and record system data.}

\State{\algorithmicuntil \hspace{0.1cm} Task learned.}
\State{\algorithmicreturn{\hspace{0.1cm}  Optimal policy $\pi^{*}_{\varphi}$.}}
\end{algorithmic} \label{Algo:1}
\end{algorithm}

The proposed framework employs Algorithm 1, to find the optimal policy $\pi^{*}_\varphi$. Now, the task is to transfer this learned policy $\pi^{*}_{\varphi}$ to a more straightforwardly interpretable and implementable PID controller. 

The proposed PID design method in the paper exploits the forward Kullback-Leibler divergence (KLD) as a loss metric to map $\pi^*_{\varphi}$ to the PID architecture in the form of controller gains. The KLD is defined as
\begin{equation}
    \mathbb{D}(f(\mathcal{X})||g(\mathcal{X}))=\sum_{\mathcal{X}} f(\mathcal{X}) \log\left ( \frac{f(\mathcal{X})}{g(\mathcal{X})} \right),
\end{equation}
where, $f(\cdot)$ and $g(\cdot)$ are pdfs over some domain $\mathcal{X}$. \AY{Intuitively, $\mathbb{D}(f(\mathcal{X})||g(\mathcal{X}))$ is a measure of the proximity of the pairs of pdfs, $f(\mathcal{X})$ and $g(\mathcal{X})$.}

Through multiple roll-outs of policy $\pi^{*}_{\varphi}$ over the system model defined in \eqref{model} the system state and control input trajectory dataset is generated. The generated dataset is used as a ground truth in the forward KLD optimization.

\section{Main Results} \label{main}
The optimal control policy $\pi^{*}_{\varphi}$ learned through the PILCO algorithm is used to generate state and control trajectory data with multiple roll-outs. The generated data is represented as \vspace{-0.3cm}
\begin{equation}\label{Dataset}
    \mathbf{E}=(\left\{\mathbf{X},\mathbf{U}\right\}),\vspace{-0.2cm}
\end{equation}
where, $\mathbf{X}\in \mathbb{R}^{M\times n}$ is the state trajectory data matrix with $M$ as system state dimension with $n$ samples. Similarly, $\mathbf{U}\in \mathbb{R}^{N \times n}$ with $N$ as the dimension of control inputs.

The collected data $\mathbf{E}$ is then augmented with the corresponding error data-set $\mathcal{E}$ is formulated as
\begin{equation}\label{error_dynm}
\mathcal{E}=\left[\tilde{e}_{1}; \dots; \tilde{e}_{n}\right],
 \mbox{with}\; \tilde{e}_i = \left(e_{\tau},\;\Delta T \sum^{T}_{\tau=0}e_{\tau},\:\frac{e_{\tau}-e_{\tau-1}}{\Delta T}\right ),
\end{equation}
where $\Delta T$ is system's sampling time. \\
Augmenting $\mathcal{E}$ from \eqref{error_dynm} to $\mathbf{E}$ in \eqref{Dataset} gives the augmented dataset $\tilde{\mathbf{E}}$, denoted by\vspace{-0.2cm}
\begin{equation}\label{augmented Data}
    \tilde{\mathbf{E}}=(\{\mathbf{E},\mathcal{E}\}).\vspace{-0.3cm}
\end{equation}\vspace{-0.2cm}

After collecting trajectory data of system states and input action for multiple different initial conditions, a conditional probability distribution can be formulated as~$p(x_{\tau}\mid x_{\tau-1},u_{\tau})$,
where $p(\cdot\mid\cdot)$ represents the conditional pdf. 

The PID controller can be tuned by minimizing the KLD between the joint pdf $P^{\pi}$ of PILCO policy $\pi^{*}_{\varphi}$ with corresponding state distribution, and the joint pdf $Q^{\phi}$ of the $e_\tau$ with PID controller pdf $c(\cdot|\cdot)$ as follows: \vspace{-0.3cm}
\begin{equation}
    \mathbb{D}(P^{\pi}||Q^{\phi})=\sum_{\tilde{\mathbf{E}}} P^{\pi}(\tilde{\mathbf{E}})\log\left ( \frac{P^{\pi}(\tilde{\mathbf{E}})}{Q^{\phi}(\tilde{\mathbf{E}})} \right),\vspace{-0.5cm}
\end{equation}
where, \vspace{-0.2cm}
\begin{equation}\label{PILCO close-loop pdf}
P^{\pi}(\tilde{\mathbf{E}}) = \prod_{\tau=0}^{T}p^{\pi}(\tilde{e}_{\tau}\mid u_{\tau},\tilde{e}_{\tau-1})c^{\pi}(u_{\tau}|\tilde{e}_{\tau-1}),
\end{equation}
is the close-loop system joint pdf for the PILCO policy $\pi^{*}_{\varphi}$, and\vspace{-0.5cm}
\begin{equation}\label{PID close-loop pdf}
Q^{\phi}(\tilde{\mathbf{E}}) = \prod_{\tau=0}^{T}p(\tilde{e}_{\tau}\mid u_{\tau},\tilde{e}_{\tau-1})c^{\phi}(u_{\tau}|\tilde{e}_{\tau-1}),\vspace{-0.1cm}
\end{equation}
is the close-loop system joint pdf with the PID controller in the feedback loop.
The PID controller pdf $c^{\phi}(u_{\tau}|\tilde{e}_{\tau-1})=\mathcal{N}(u^{\phi}_{\tau},\sigma_{\phi})$, where the mean $u^{\phi}_{\tau}$ and covariance $\sigma_{\phi}$ parameterized by PID gains~$\phi$.

\begin{assump}\label{Assumption: Convex}
 We assume that the distribution $Q^{\phi}(\mathbf{E})$ is of exponential nature, as it allows $\mathbb{D}(\cdot||\cdot)$ to be convex in $\phi$.
\end{assump}

Based on the above assumption, we present the following lemma to map the PILCO policy into optimal PID gains. 

\begin{lem}\label{lem: KLD lemma}
Given a tuned RL policy $\pi^{*}_{\varphi}$, a set of optimal PID parameters can be obtained if $\argmin_{\phi} \mathbb{D}(\cdot||\cdot)<\epsilon$, where~$\epsilon$ is some constant with positive real value and $\phi$ is the set of PID parameters $\{K_{P},K_{I},K_{D}\}$.
\end{lem}
\begin{proof}
For the system in~\eqref{model} and controller pdfs in~\eqref{PILCO close-loop pdf}, \eqref{PID close-loop pdf} the KLD can be written in the additive form as
\begin{equation*}\label{KLD_ADD}
\begin{split}
& \mathbb{D}(P^{\pi}||Q^{\phi})=
=\sum_{\tilde{\mathbf{E}}}P^{\pi}(\tilde{\mathbf{E}})\log\left(\frac{P^{\pi}(\tilde{\mathbf{E}})}{Q^{\phi}(\tilde{\mathbf{E}})}\right)\\
& =\sum_{\tilde{\mathbf{E}}}P^{\pi}(\tilde{\mathbf{E}})\left(\log(P^{\pi} (\tilde{\mathbf{E}}))-\log(Q^{\phi}(\tilde{\mathbf{E}}))\right)
\end{split}
\end{equation*}
\begin{equation*}
=\sum_{\tilde{\mathbf{E}}}\underbrace{P^{\pi}(\tilde{\mathbf{E}})\log(P^{\pi} (\tilde{\mathbf{E}}))}_{\rm{entropy}}-\sum_{\tilde{\mathbf{E}}}\underbrace{P^{\pi}(\tilde{\mathbf{E}})\log(Q^{\phi}(\tilde{\mathbf{E}}))}_{\rm{cross-entropy}}.
\end{equation*}
The first term is entropy $(\mathcal{H}(\cdot))$ which doesn't depend on $\phi$. So, we only focus on the second term which is cross-entropy, derived as
\begin{equation}
\begin{split}
    &\sum_{\tilde{\mathbf{E}}}P^{\pi}(\tilde{\mathbf{E}})\log(Q^{\phi}(\tilde{\mathbf{E}}))=\frac{1}{Z_{P}}\sum_{\tilde{\mathbf{E}}}P^{\pi}(\tilde{\mathbf{E}})\log\left(\frac{Q^{\phi}(\tilde{\mathbf{E}})}{Z_{Q}}\right)\\
    &=\frac{1}{Z_{P}}\sum_{\tilde{\mathbf{E}}}\bar{P}^{\pi}(\tilde{\mathbf{E}})\log(\bar{Q}^{\phi}(\tilde{\mathbf{E}}))-\frac{1}{Z_{P}}\sum_{\tilde{\mathbf{E}}}\bar{P}^{\pi}(\tilde{\mathbf{E}})\log(Z_{Q})\\
    &=\frac{1}{Z_{P}}\sum_{\tilde{\mathbf{E}}}\bar{P}^{\pi}(\tilde{\mathbf{E}})\log(\bar{Q}^{\phi}(\tilde{\mathbf{E}}))-\log(Z_{Q})\left(\frac{1}{Z_{P}}\sum_{\tilde{\mathbf{E}}}\bar{P}^{\pi}(\tilde{\mathbf{E}})\right)\\
    &=\frac{1}{Z_{P}}\sum_{\tilde{\mathbf{E}}}\bar{P}^{\pi}(\tilde{\mathbf{E}})\log(\bar{Q}^{\phi}(\tilde{\mathbf{E}}))-\log(Z_{Q}),
\end{split}
\end{equation}
where data normalization is performed using the normalizing constants $Z_{P}$ and $Z_{Q}$. 
The gradient for the optimization problem when $Q^{\phi}$ belongs to the exponential family can derived as
\begin{equation}\label{GRAD}
    \begin{split}
        &\nabla\left[\frac{1}{Z_{P}}\sum_{\tilde{\mathbf{E}}}\bar{P}^{\pi}(\tilde{\mathbf{E}})\log(\bar{Q}^{\phi}(\tilde{\mathbf{E}}))-\log(Z_{Q})\right]\\
        & =\frac{1}{Z_{P}}\sum_{\tilde{\mathbf{E}}}\bar{P}^{\pi}(\tilde{\mathbf{E}})\nabla[\log(\bar{Q}^{\phi}(\tilde{\mathbf{E}}))]-\nabla\log(Z_{Q}).
    \end{split}
\end{equation}

The optimization problem can be formulated w.r.t $\phi$ as
\begin{equation}\label{Optimization}
\begin{split}
    &\arg\min_{\phi}\mathbb{D }\left(P^{\pi}\mid\mid Q^{\phi}\right) :=\\ 
    &=\arg\min_{\phi}\mathbb{E}_{\tilde{\mathbf{E}}\sim P}\left[-\log(Q^{\phi}(\tilde{\mathbf{E}}))\right] +\mathcal{H}(P^{\pi} (\tilde{\mathbf{E}}))\\
    & =\arg\min_{\phi}\mathbb{E}_{\tilde{\mathbf{E}}\sim P}\left[-\log(Q^{\phi}(\tilde{\mathbf{E}}))\right]\\
    &=\arg\max_{\phi}\mathbb{E}_{\tilde{\mathbf{E}}\sim P}\left[\log(Q^{\phi}(\tilde{\mathbf{E}}))\right].
\end{split}
\end{equation}
Notice that this is identical to the maximum likelihood estimation objective. In other words, the objective~\eqref{Optimization} will sample points from $P^{\pi}(\tilde{\mathbf{E}})$ and try to maximize the probability of the points under $Q^{\phi}(\tilde{\mathbf{E}})$. 
A good approximation under the forward KLD objective thus satisfies wherever $P(\cdot)$ has high probability $Q(\cdot)$ must also have high probability.
Consider this mean-seeking behavior, because the approximate distribution $Q(\cdot)$ must cover all the modes and regions of high probability in $P(\cdot)$. It is worth to note that the forward KLD doesn't penalize $Q(\cdot)$ for having high probability mass where $P(\cdot)$ does not.\\
Being convex, the objective function eventually converges to a small neighborhood of zero $(\epsilon)$. Thus, minimization of $\mathbb{D}$ over $\phi$ yields a set of optimal PID gains $\phi^*$. This completes the proof.
\end{proof}
\begin{rem}\label{rem:robustness}
It is worth highlighting that, while learning the PILCO policy $\pi^{*}_{\varphi}$, plant noise and system parameter uncertainty are introduced in each roll-out, thereby yielding a robust PILCO policy. During the design of optimal PID tuning parameters by Lemma~\ref{lem: KLD lemma}, the robustness of $\pi^{*}_{\varphi}$ is inherited to the $\phi^{*}$, which leads to a robust PID controller for a system parameters ($\psi$) distribution $\mathcal{N}(\mu_{\psi},\sigma_{\psi})$.
\end{rem}
We present the following algorithm to summarize the designing of a set of robust PID tuning parameters.
~\begin{algorithm}[h]\label{Algo2}
\caption{Learning $\phi^*$ through $\pi^*_\varphi$}
\begin{algorithmic}[1]
\State{\textbf{Input:}\quad $\pi^{*}_{\varphi}$ from Algorithm \ref{Algo:1} }
\State{\textbf{Step 1:} \begin{itemize}
    \item Roll-out learned $\pi^*_\varphi$ to form $\mathbf{E}$ as in \eqref{Dataset}.
    \item Formulate $\mathcal{E}$ as per \eqref{error_dynm}.
    \item Data augmentation as per \eqref{augmented Data}.
    \item Initialize PID parameters $\phi$ as random variables.
\end{itemize}
}
\State{\textbf{Step 2:}}\begin{itemize}
    \item According to the Lemma \ref{lem: KLD lemma},
    
    \algorithmicrepeat

\begin{itemize}
    \item Formulate optimization problem as per \eqref{Optimization} over the domain of $\tilde{\mathbf{E}}$.
    \item Minimize $\mathbb{D}(\cdot||\cdot)$ over $\phi$ in \eqref{Optimization}.
    \item Update $\phi$ in \eqref{Optimization} and \eqref{GRAD}.

\end{itemize}

\algorithmicuntil \quad \eqref{Optimization} converges to $\epsilon$ \end{itemize}

\State{\algorithmicreturn{\quad $\phi^*$}}
\end{algorithmic} \label{Algo:2}
\end{algorithm}

\section{Simulation Results and Analysis}

The proposed method is employed to devise robust PID controller gains for a benchmark underactuated cart-pole system presented in~\cite{olfati1999fixed}. It is defined by four states $s=\{\mathrm{x},\,\dot{\mathrm{x}},\,\theta_{\mathrm{rad}},\,\dot{\theta}_{\mathrm{rad}}\}$, where $\mathrm{x}$ is the cart-position, $\dot{\mathrm{x}}$ is the cart velocity, $\theta_{\mathrm{rad}}$ and $\dot{\theta}_{\mathrm{rad}}$ are pole angle and angular velocity, respectively. The desired set point is $s_{des}=\{0,0,0,0\}$, which is the upright position for the pole (unstable equilibrium). 
\AY{The control structure~\eqref{PID_1} is used for the cart-pole example. The feedback from channels for $x$ and $\theta_{\rm{rad}}$ are used to derive the control input $u = K\tilde{e}^{\top}$, where $K$ and $\tilde{e}$ are defined in~\eqref{PID_1} and ~\eqref{error_dynm}, respectively.}

\subsection{Learning PID gains from $\pi^{*}_{\varphi}$}
The robust PID gains are learned from the PILCO policy $\pi^{*}_{\varphi}$ by solving~\eqref{Optimization}. 
\begin{figure*}[h!]
  \centering
  \begin{tabular}[b]{c}
    \includegraphics[width=.33\textwidth]{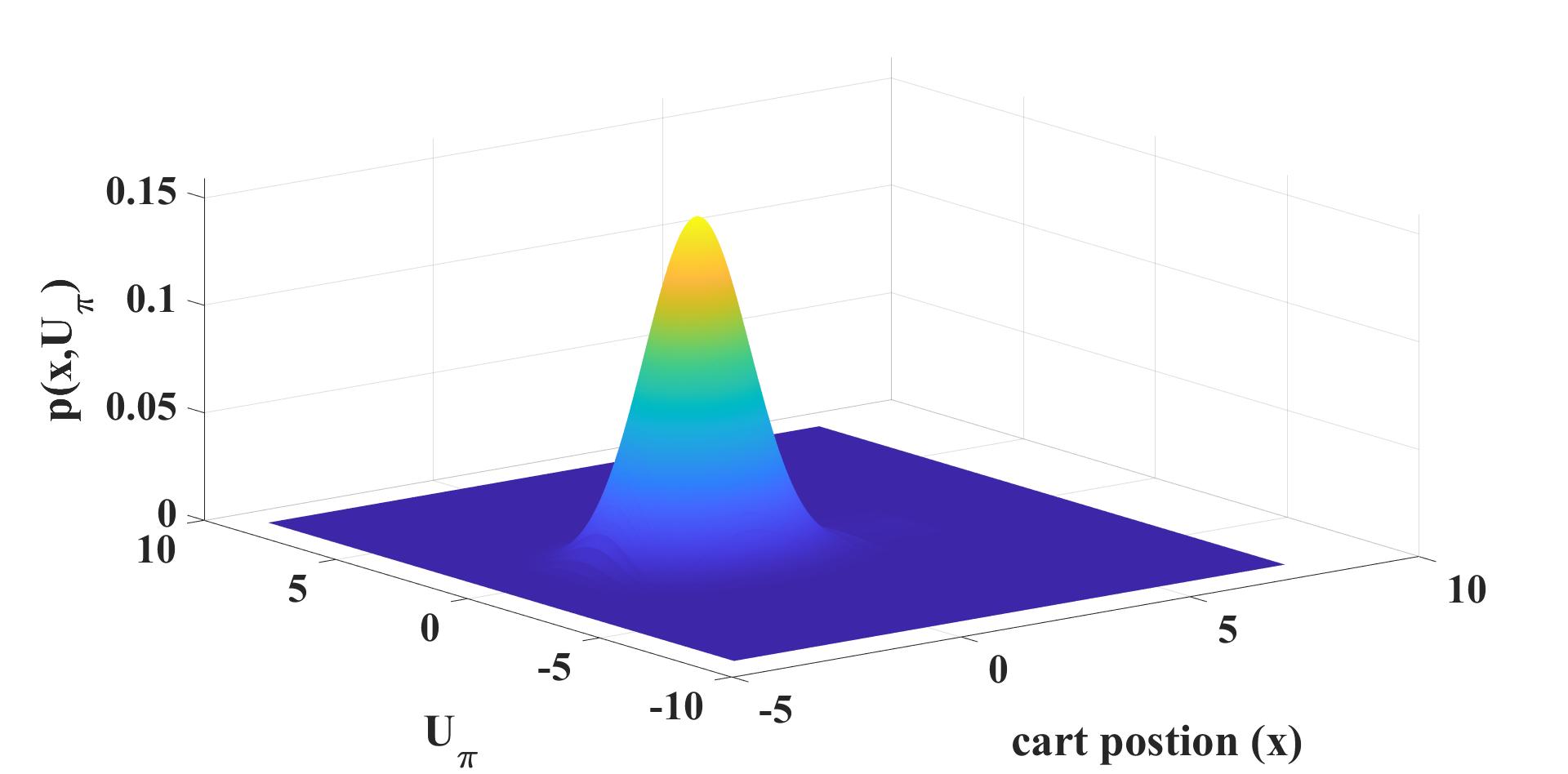} \\
    \small (a)
  \end{tabular} 
  \hspace{-1cm}
  \begin{tabular}[b]{c}
    \includegraphics[width=.33\textwidth]{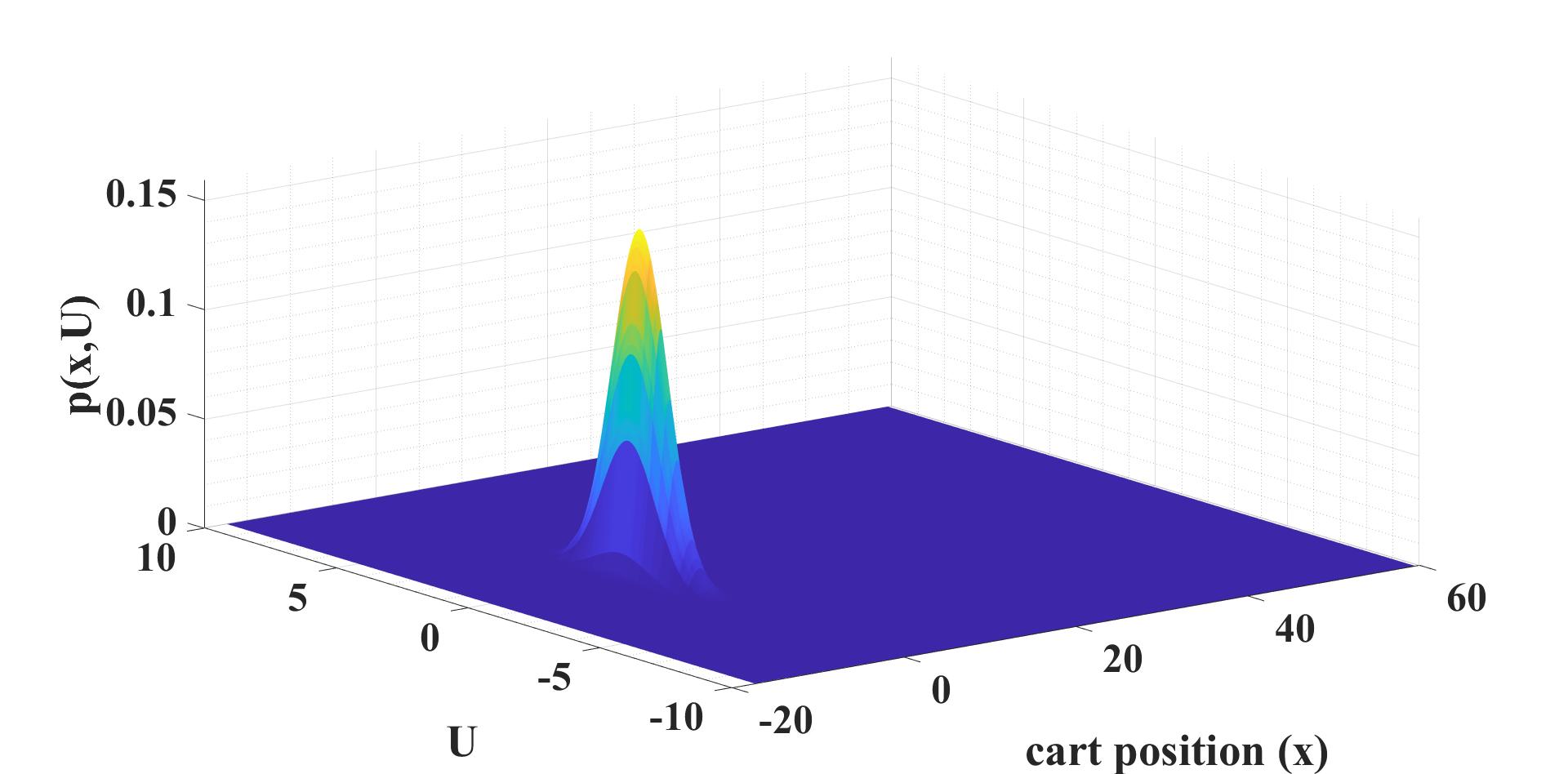} \\
    \small (b)
  \end{tabular}
  \hspace{-1cm}
  \begin{tabular}[b]{c}
    \includegraphics[width=.33\textwidth]{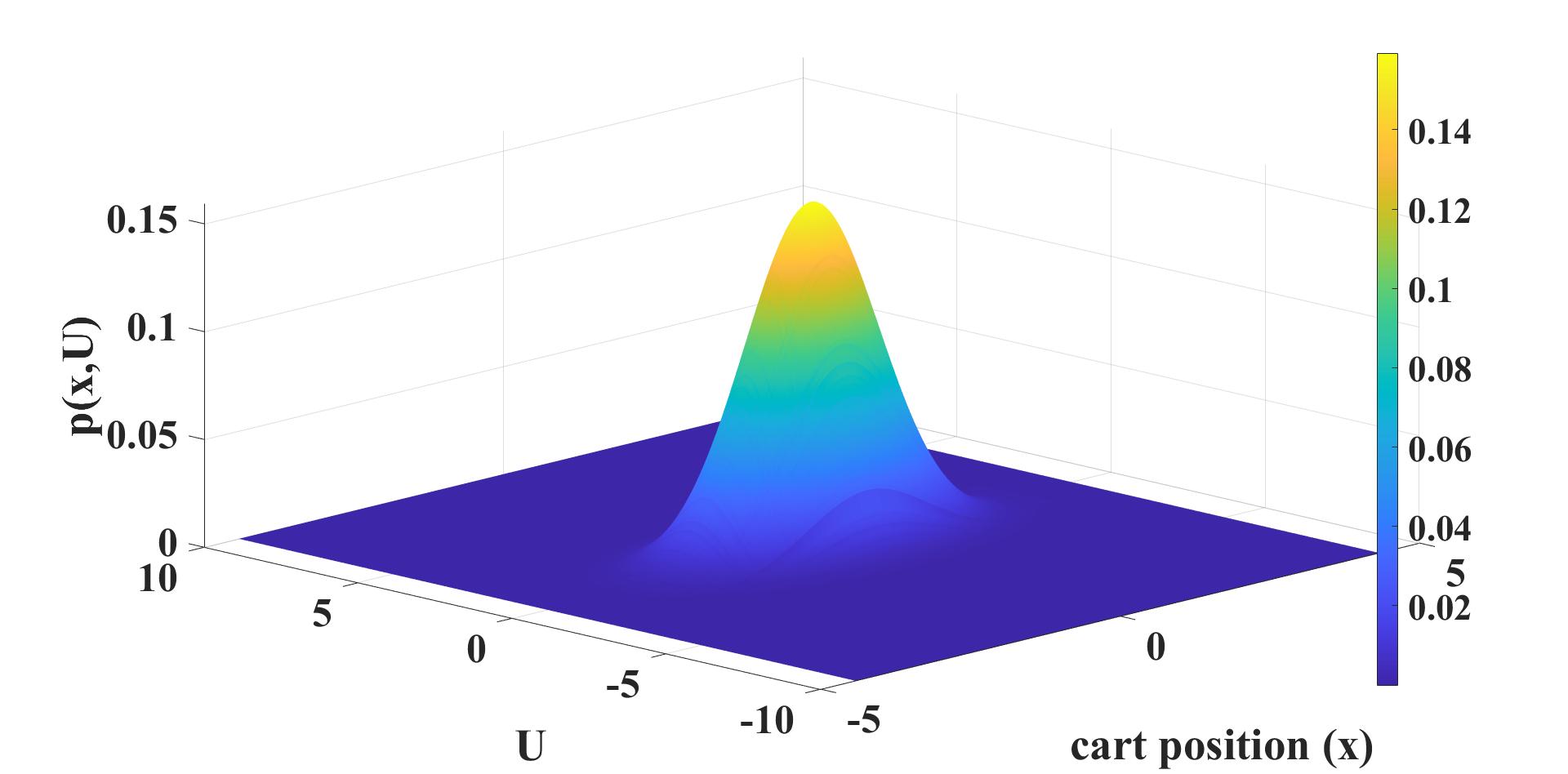} \\
    \small (c)
  \end{tabular}
  \hspace{-1cm}
  \begin{tabular}[b]{c}
    \includegraphics[width=.33\textwidth]{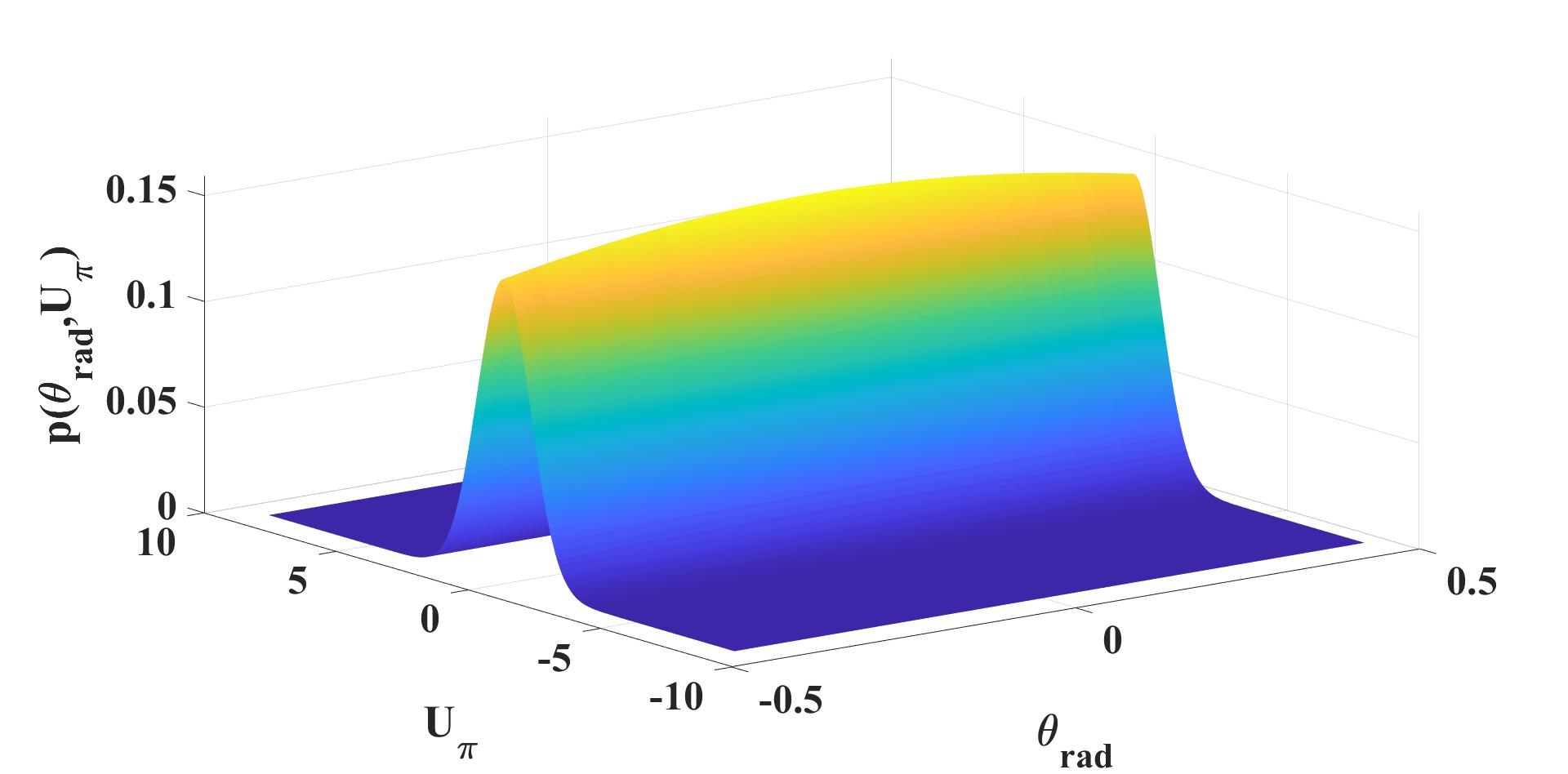} \\
    \small (d)
  \end{tabular}
  \hspace{-1cm}
  \begin{tabular}[b]{c}
    \includegraphics[width=.33\textwidth]{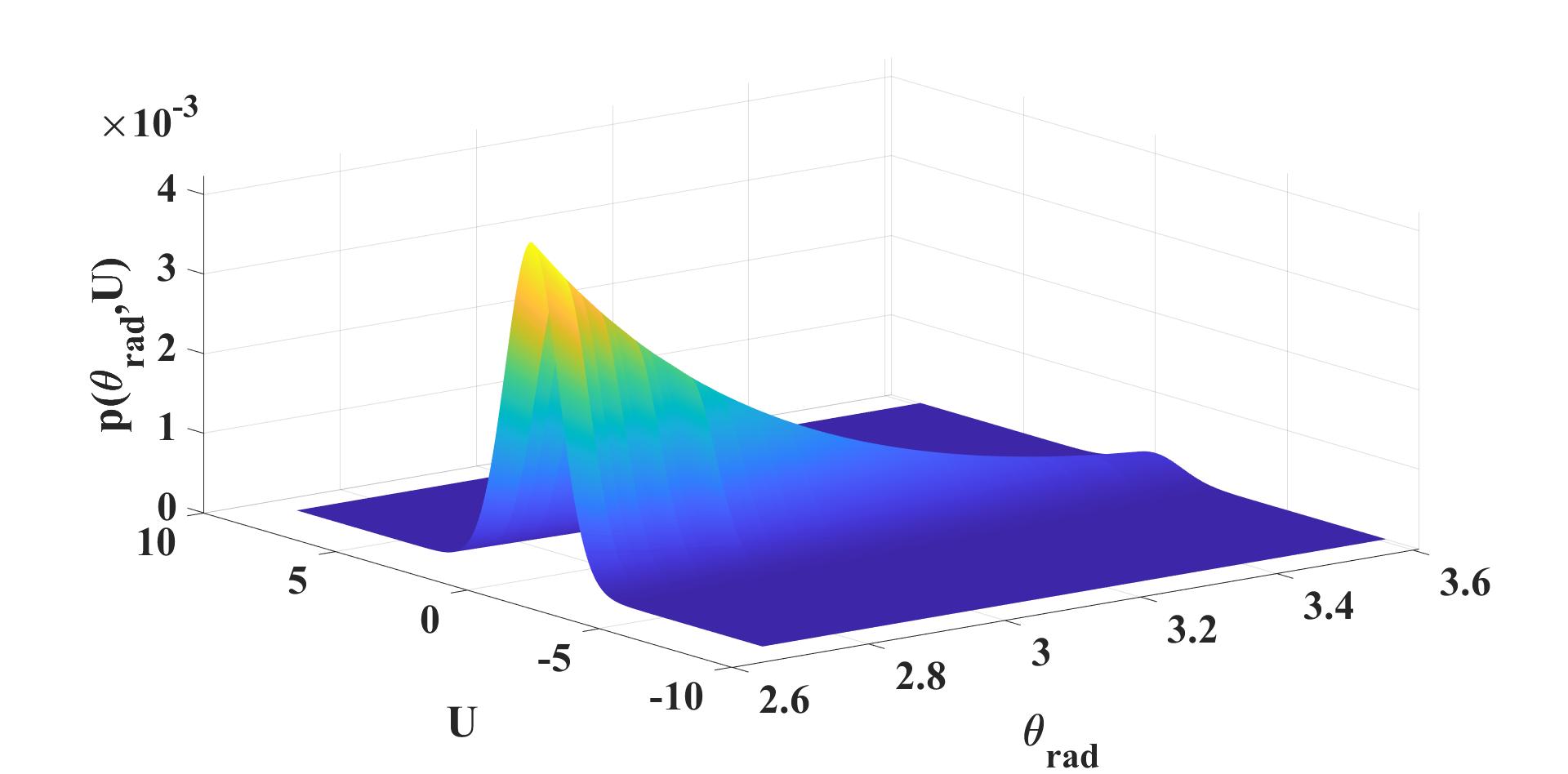} \\
    \small (e)
  \end{tabular}
  \hspace{-1cm}
  \begin{tabular}[b]{c}
    \includegraphics[width=.33\textwidth]{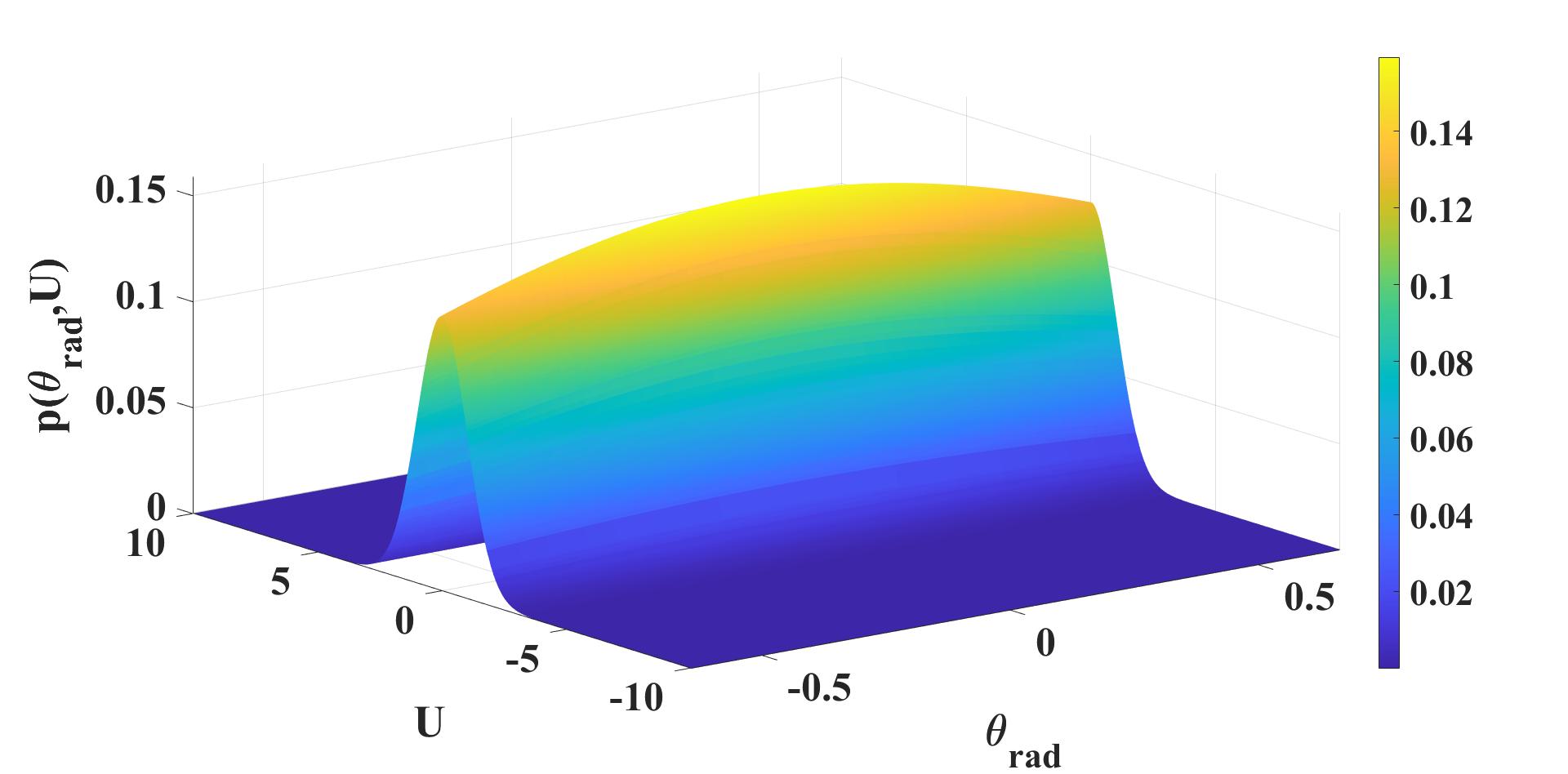} \\
    \small (f)
  \end{tabular}
 
\caption{Minimizing $\mathbb{D}(\cdot||\cdot)$, (a)~joint pdf of $\pi^{*}_{\varphi}$ and cart position $(\mathrm{x})$; (b)~initial joint pdf of PID controller and $\mathrm{x}$; (c)~joint pdf of PID controller and $\mathrm{x}$ after convergence to $\phi^{*}$; (d)~joint pdf of $\pi^{*}_{\varphi}$ and pole angle $(\theta_{rad})$; (e)~initial joint pdf of PID controller and $\theta_{rad}$; (f)~joint pdf of PID controller and $\theta_{rad}$ after convergence to $\phi^{*}$.}\label{joint_pdf_plot}
\end{figure*}
By following Lemma~\ref{lem: KLD lemma}, the KLD analysis is depicted in Fig.~\ref{joint_pdf_plot}, where Figs.~\ref{joint_pdf_plot}(a) and \ref{joint_pdf_plot}(d) show the joint pdf of $\pi^{*}_{\varphi}$ with $\mathrm{x}$ and $\theta_{\mathrm{rad}}$, respectively. The initial joint pdf of the PID controller with $\mathrm{x}$ and $\theta_{\mathrm{rad}}$ is shown in Figs.~\ref{joint_pdf_plot}(b) and \ref{joint_pdf_plot}(e), respectively, where initial PID parameters are randomly generated from a distribution. It can be seen in Figs.~\ref{joint_pdf_plot}(c) and \ref{joint_pdf_plot}(f) that at the $900^{th}$ iteration, joint pdfs of the PID and $\pi^{*}_{\varphi}$ follow Lemma~\ref{lem: KLD lemma}.

\subsection{Stability analysis of the closed-loop system}
To provide a certificate of stability for PID controlled close-loop system, roll-outs with the boundary points of the region of attraction (ROA) as initial conditions are performed. The ROA of the close-loop system is estimated using the stable trajectory data collected through roll-outs. The method for ROA estimation derived in \cite{colbert2018using} depends on the system trajectory data instead of the nonlinear ordinary differential equations. The method doesn't rely on the analytically determined Lyapunov function $V(x_{\tau})$, it uses the converse Lyapunov theorems defined as, $V(x_{\tau})=\int_{0}^\infty \left | \right | f(x_{\tau})\left | \right |^2 d\tau, $
where $f(x_{\tau}^i)$ for $i=\{1,n\}$ is the system trajectory over $n$ initial conditions.  After obtaining the optimal Lyapunov function $V^{*}(x_\tau)$, the ROA is estimated as a maximal level set of the $V^{*}$.~\begin{figure}[h!]
    \centering
    \includegraphics[width=0.4\textwidth, height=0.45\linewidth]{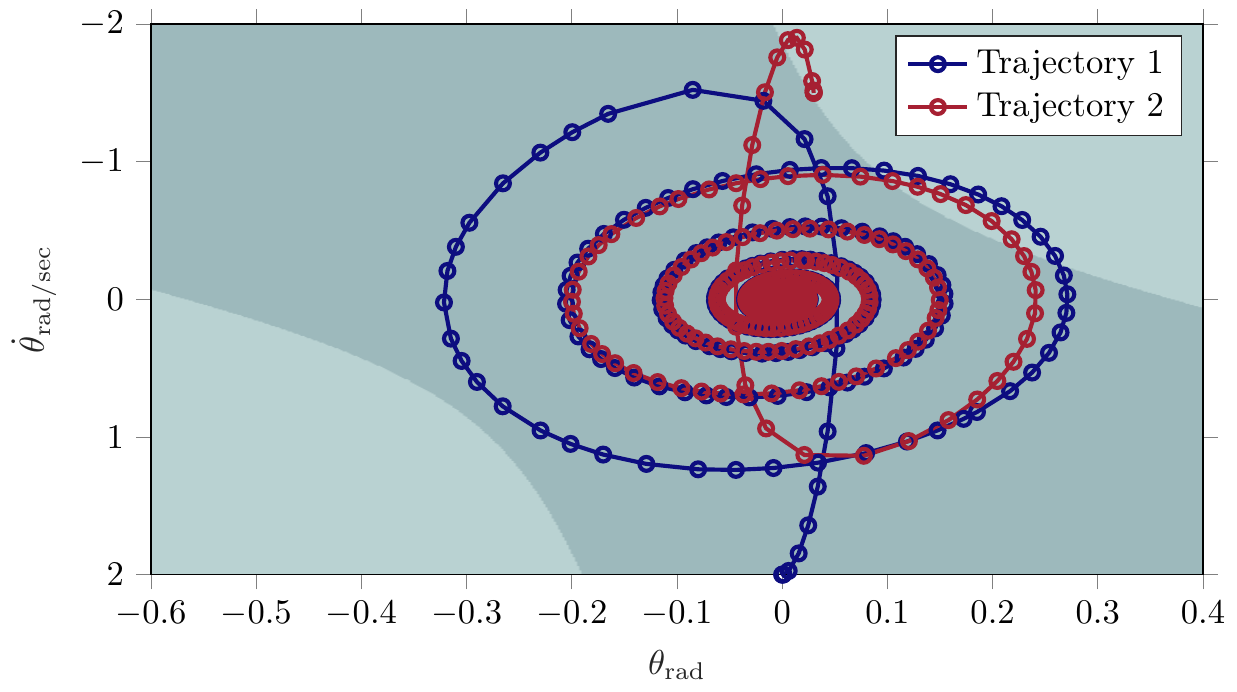}
    \caption{ROA \AY{(dark shaded region)} of the cart-pole system for an unstable equilibrium and the converging trajectories.}
    \label{ROA_PLOT}
\end{figure}

ROA of the cart-pole system around the unstable equilibrium point
$\{\theta,\dot{\theta}\}=\{0,0\}$ is shown in the Fig.~\ref{ROA_PLOT}, wherein the system is controlled by PIDs designed using PILCO. The PID controlled trajectories with boundary initial conditions, converging to the unstable equilibrium are also plotted in the Fig.~\ref{ROA_PLOT}. This shows that the controller derived using Algorithm~\ref{Algo:2} successfully stabilizes the system to unstable equilibrium even with ROA boundary initial conditions.

\subsection{Controller performance under disturbance and system parameter uncertainty}
\begin{figure}[h!]
\centering
\input{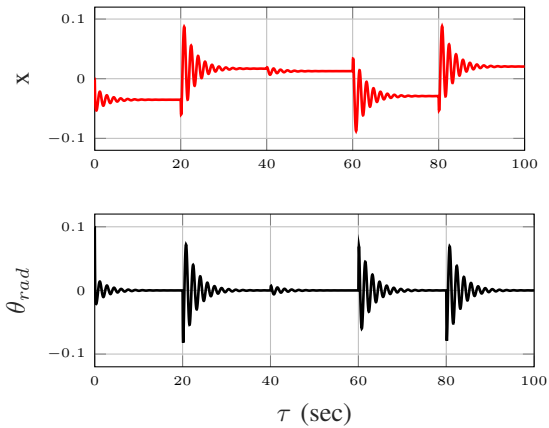}
\caption{System trajectory for matched disturbance on $\mathrm{x}$.}\label{matched response}
\end{figure}

\begin{figure}[h!]
\centering
\input{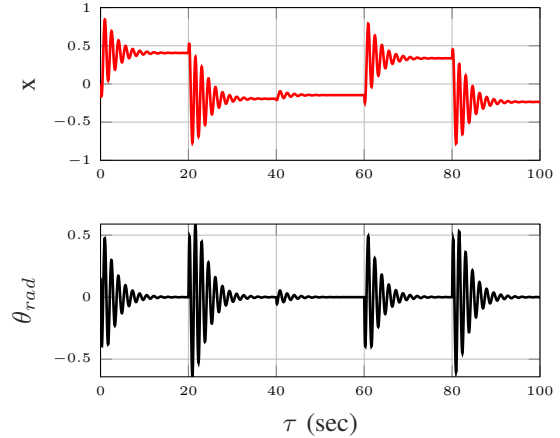}
\caption{System trajectory for unmatched disturbance on $\theta$.}\label{unmatched response}
\end{figure}

When a disturbance on the actuated channel is applied it is known as matched disturbance, and a disturbance on the unactuated channel is a unmatched disturbance. The performance analysis of the tuned PD controller using proposed method under both matched and unmatched disturbances is provided. The close-loop system response for matched disturbance on the cart is shown in Fig.~\ref{matched response}, while Fig.~\ref{unmatched response} shows system response for unmatched disturbance. It is clear from the Fig.~\ref{matched response} and Fig.~\ref{unmatched response} that controller can render system asymptotically stable for both kind of disturbances. The controller has tolerance for noise in both PIDs' feedback channel. Thus, the conclusion can be drawn that the learned controller gains are less prone to the matched and unmatched noise (which enters the system through input).

For a controller to be called robust it should stabilize the system under model/parameter uncertainty for some given range. In Algorithm~\ref{Algo:1}, plant noise and uncertainty in data is introduced to yield a robust policy $\pi^{*}_{\varphi}$, this property is transferred to the PID gains obtained through Algorithm~\ref{Algo:2}. Fig.~\ref{parameter uncertainty} depicts close-loop system response for parameter uncertainty. \AY{System parameters, namely pendulum mass $m$ (kg) and pole length $l$ (m) are sampled from a distribution $\mathcal{N}(\mu_{m,l},\Sigma_{m,l})$, where the parameter mean $\mu_{m,l}=[0.2;0.5]$ and variance $\Sigma_{m,l}= I_{2\times2}[0.0025,0.005]^\top$. The figure clearly shows the robustness of the designed controller. Finally, we replaced PILCO with the MBPO algorithm to design the PID controllers for the cart-pole system. Fig.~\ref{MBPO-PID} shows the closed-loop performance, wherein the system is stabilized at $s_{des}=\{0,0,0,0\}$.
\begin{figure}[h!]
\centering
\input{Parameter_Uncertainity}
\end{figure}
\begin{figure}[h!]
    \centering
    \includegraphics[width=0.46\textwidth, height=0.45\linewidth]{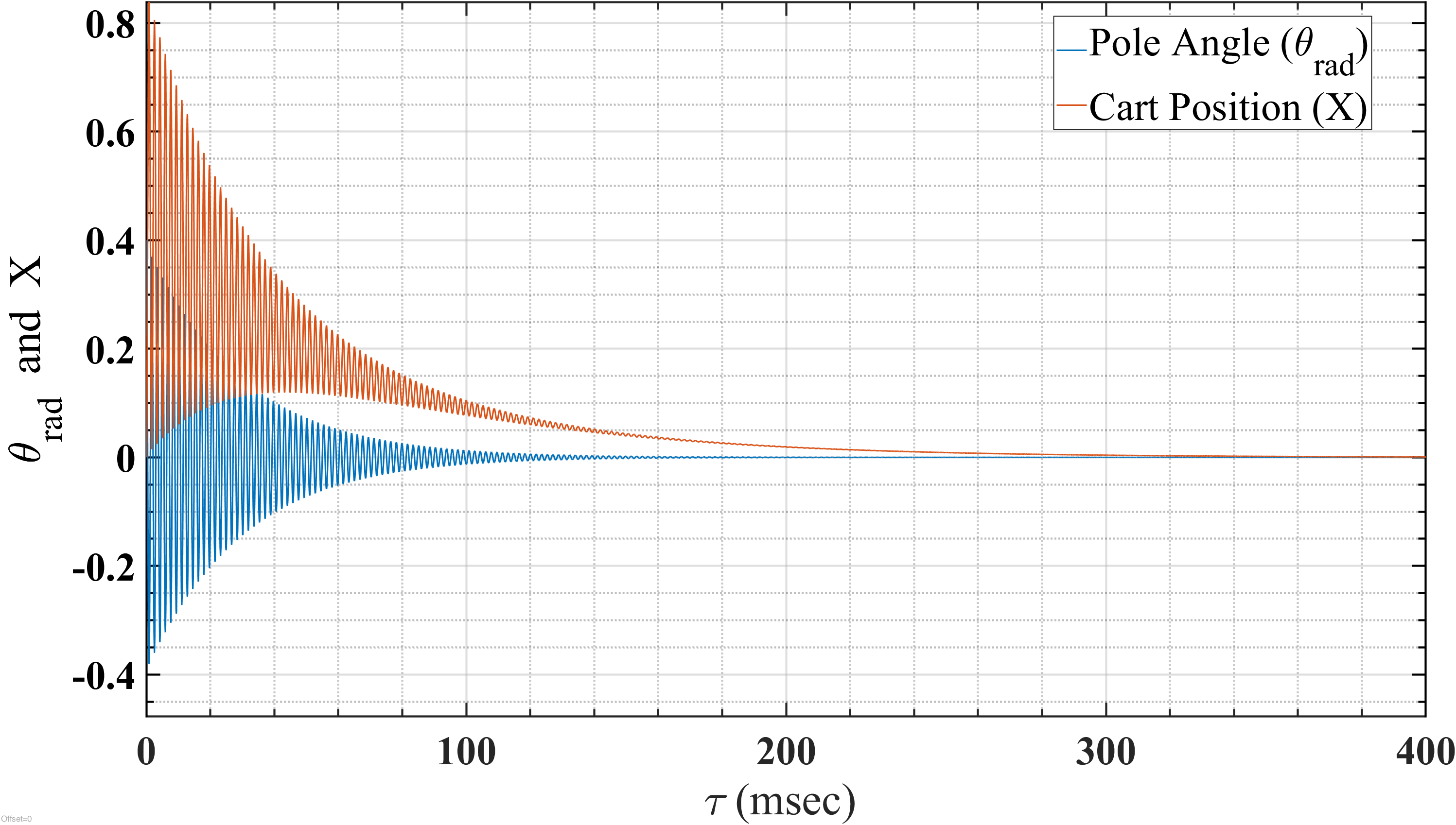}
    \caption{MBPO-PID performance on cart-pole.}
    \label{MBPO-PID}
\end{figure}
}

\section{Conclusion}\label{Conclusion}
\AY{A generalized framework for designing PID controllers has been presented. In particular, based on the KLD and leveraging the state trajectory data and control policy of data-efficient model-based RL methods, a more interpretable PID control design technique has been presented. Further, PIDs have been designed using PILCO and MBPO algorithms, and closed-loop behavior has been demonstrated. Robust performance has been observed under various disturbances and parameter uncertainty. }



\bibliography{CDC22_LCSS.bib}

\end{document}